# On tit for tat: Franceschini and Maisano versus ANVUR regarding the Italian research assessment exercise VQR 2011-2014[1]

*Giovanni Abramo[a], Ciriaco Andrea D'Angelo[b,a]*

[a] Laboratory for Studies in Research Evaluation
Institute for System Analysis and Computer Science (IASI-CNR)
National Research Council of Italy

[b] Department of Engineering and Management
University of Rome "Tor Vergata"

## 1. Introduction

The response by Benedetto, Checchi, Graziosi & Malgarini (2017) (hereafter "BCG&M"), past and current members of the Italian Agency for Evaluation of University and Research Systems (ANVUR), to Franceschini and Maisano's ("F&M") article (2017), inevitably draws us into the debate. BCG&M in fact complain "that almost all criticisms to the evaluation procedures adopted in the two Italian research assessments VQR 2004-2010 and 2011-2014 limit themselves to criticize the procedures without proposing anything new and more apt to the scope". Since it is us who raised most criticisms in the literature, we welcome this opportunity to retrace our vainly "constructive" recommendations, made with the hope of contributing to assessments of the Italian research system more in line with the state of the art in scientometrics. We see it as equally interesting to confront the problem of the failure of knowledge transfer from R&D (scholars) to engineering and production (ANVUR's practitioners) in the Italian VQRs. We will provide a few notes to help the reader understand the context for this failure. We hope that these, together with our more specific comments, will also assist in communicating the reasons for the level of scientometric competence expressed in BCG&M's heated response to F&M's criticism.

## 2. ANVUR

ANVUR began operations in May 2011, with the appointment of seven full-time members of the executive committee, following public competition for the positions (full disclosure: one of us applied unsuccessfully to the original 2010 call, as well as the 2015

---

call for renewal of committee members).[2] All members of the 2011 committee were professors (one each in physics, engineering, medicine, veterinary sciences, sociology, two in economic sciences). At that time, only one (Andrea Bonaccorsi) had ever authored a Scopus or WoS indexed publication on research evaluation. Professor Bonaccorsi, an economist by education, showed marginal diversification of his scientific production in research evaluation. A managing director was hired soon after the executive committee assumed operations, followed by several staff members and collaborators on temporary contracts. On 7 November 2011, ANVUR, acting on authority of the Italian Ministry of Education, University and Research (MIUR), launched the 2004-2010 VQR. Unfortunately, five months from appointment would likely have been too little for adequate conception, design and organization of such a large scale exercise, even for an executive committee composed of seven of the world's most experienced scientometricians. For the first VQR, Sergio Benedetto (one of the authors of the response to F&M), a professor in telecommunications engineering, was appointed as VQR Coordinator, with Andrea Bonaccorsi as Assistant Coordinator. When the second VQR (2011-2014) was launched (3 September 2015) the executive committee had for the most part been newly appointed, with only Sergio Fantoni, a physicist (president of the original committee), and Sergio Benedetto held over. All five new entries were again professors: Andrea Graziosi (an author of the response to F&M), professor in contemporary history, succeeded Sergio Fantoni as Committee President; Daniele Checchi (also a responder to F&M), a professor in economics, succeeded Sergio Benedetto as VQR Coordinator. The remaining members of the current committee are a neuroscientist, a mathematician, and a surgeon. None of the new entries shows a publication on research evaluation in indexed journals. Finally, the last author of the response to F&M, Marco Malgarini, joined ANVUR staff in 2012 and is now head of the research evaluation area. At the time of hiring he had no indexed publications on research evaluation, but now counts three.

## 3. Responses to the specific comments by BCG&M

Before dealing with the "constructive" criticism issue raised by BCG&M, we would like to comment on their specific responses to F&M.

In Section 2, paragraph 7 of their response, BCG&M mention two publications as heavily critical to the 2004-2010 VQR. Because these are by Italian authors, BCG&M dispute F&M's consideration that such criticism is "international", with the inference that this also makes it of little significance. Our view is that science has no nationality, and that good science does not depend on the authors' location, rather on the truths that it hopefully embeds. Perhaps F&M appreciated the "international" standing of the criticism in the sense of the global context of the hosting journals. (There have also been more than two critical

---

[2] The decree founding ANVUR stipulated a maximum office of five years for two members of the original executive committee, four years for three members, and three years for the other two members. For subsequent appointments, office is to be four years maximum.

articles by Italian authors, all published in international bibliometric journals, as noted in Section 5.)

In Section 3.1, BCG&M state: "To show the inconsistency of this exercise suffices to say that the VQR 2011-2014 results published on February 22$^{nd}$, 2017, show that only 32.6% of the proposed products were assessed as excellent, and a total of 63.4 are in the two classes A or B." We do not enter into the "triviality" of F&M's critique, as labeled by BCG&M. Whether trivial or not, the reasoning offered by BCG&M is not at all sufficient as a response. In fact, BCG&M forgo responding to the demonstrated inefficiency in the selection of best publications (in keeping with VQR bibliometric criteria) to be submitted for evaluation. Abramo, D'Angelo and Di Costa (2014), showed that in the 2004-2010 VQR, universities' inefficient selection had worsened the maximum scores achievable by up to 23% to 32%.

BCG&M further object to F&M's recommendation "to evaluate all the products published in the VQR period, at least for the bibliometric scientific areas" because "this is contrary to the principle of treating all areas equally and, by the way, has never been adopted in RAEs or REF assessments in the United Kingdom". In fact, it is exactly through demanding the same number of products per researcher for all areas that ANVUR contravenes the principle of treating all areas equally, therefore favoring certain entities over others. We do not need to be bibliometricians to realize that intensity of publication varies across fields.[3] Three research products over four years could well be much more than the output possible for an average historian, but far less than that of an average experimental physicist. Research institutions with a higher share of professors in low publication-intensive fields are then disfavored. By further inference, "treating all areas equally" should mean applying the same evaluation method (either peer-review or bibliometrics) to all. In fact, it has been shown that in the VQR case "bibliometric scores are on average significantly higher than those obtained with the peer review" (Cicero, Malgarini, Nappi, and Peracchi, 2013). Why did ANVUR then not enforce the "equality" principle in this case, rather than offering bibliometrics for some and not others?

In Section 3.2, BCG&M object to F&M's recommendation "not to use the journal impact as one of the two bibliometric indicators as a proxy of the article impact". BCG&M explain their choice stating: "first, the VQR is evaluating significantly large communities of researchers, second, to do so, it has to evaluate relatively recent publications (as an example, articles published in 2014 based on citations received up to February 2016), and third, it uses two bibliometric indicators rather than one (using two indicators is generally more robust than using one)." We fail to understand what evaluating large or small communities has to do with the choice of the impact indicator of publications, and why "to do so" one needs "to evaluate relatively recent publications". Second, the 2011-2014 VQR indeed did apply the combination of journal impact and citations (C-J) to assess the impact of 2011-2013 publications, while recurring to peer-review for all 2014 publications except for those resulting as excellent. In most cases then, the citation window was not so impossibly short as BCG&M would have us believe. Third, if using two indicators is more robust than using one, why not use twenty? As scientists, BCG&M must know it is the inappropriate selection and combination of indicators that would distort a measure.

---
[3] Descriptive statistics of Italian publication distributions in 192 fields are available in D'Angelo and Abramo (2015).

With reference to the choice of using a C-J combination, BCG&M further note that "referring to what has been suggested and published in the past as the "Truth" has always prevented science to progress, and it represents a conservative attitude that should not belong to scientists". We are instead profoundly convinced that a scientific truth of the past is always better than a new "Truth" arrived at through non-scientific method, as was the C-J combination proposed and applied by ANVUR in the 2011-2014 VQR. In a paper published shortly before F&M's work, with the transparent title "Refrain from adopting the combination of citation and journal metrics to grade publications, as used in the Italian national research assessment exercise (VQR 2011-2014)" (Abramo & D'Angelo, 2016), we state that "any new form of measure that could better predict future impact, including a hybrid indicator, would be welcomed by the bibliometrics community". In spite of this openness, we could not help but notice that the C-J indicator proposed by ANVUR, with discretional weighting of C and J varying by discipline and year, lacked any empirical demonstration of validity as a predictor of impact. It was simply a flight of fancy by ANVUR. From this work, we also extract an example of the approach of explicit, constructive recommendation forwarded to ANVUR: "For the benefit of those responsible for future evaluations, and for the achievement of national, institutional and individual academic goals, we demonstrate that the simple citation count is a better proxy of a publication impact than the C-J metric adopted by ANVUR."

**4. Constructive criticisms**

In the following, we extract some of the constructive (but vainly offered) criticisms and recommendations forwarded to ANVUR/VQR over the years.

In 2009, two years before the launch of the 2004-2010 VQR by ANVUR, Abramo, D'Angelo, and Caprasecca (2009) analysed the results of the first Italian research assessment exercise, the 2001-2003 Triennial Research Evaluation (VTR), which was based on peer-review evaluation of a subset of scientific products. The conclusions, totally ignored by ANVUR in the subsequent VQRs, were as follows:

"Comparing to the peer review process, the bibliometric approach (enabling the evaluation of all publications) permits:
i) avoiding the weakest phase in peer review, meaning the selection of articles by individual research institutions;
ii) assessing research productivity, both in quantitative and qualitative terms;
iii) significantly reducing the costs and times for implementation."

A year later, Abramo, D'Angelo, and Viel (2010) provided evidence that the performance rankings resulting from national assessments could be quite sensitive to the size of the share of publications to be submitted for evaluation. In the same year, Abramo, D'Angelo, and Solazzi (2010), with reference to national research assessment exercises, provided a measure of the distortion of performance rankings when labor input is treated as uniform. Accounting for academic rank when comparing performance of Italian institutions, would avoid favoring those with a higher proportion of full professors, who are in fact more productive (Abramo, D'Angelo & Di Costa, 2011). In June 2011, Abramo and D'Angelo (2011) provided evidence, in bibliometric areas, of the superiority of assessment exercises based on all indexed publications over those based on a selection of them. The comparison between the two was conducted in terms of the essential parameters of any measurement system: accuracy, robustness, validity, functionality, time and costs. In July 2011, *Scientometrics* published online the work by Abramo, D'Angelo, and Di Costa (2011) measuring the university ranking distortions in research assessments based on a subset of publications.

Before formulating the first VQR, ANVUR had then enough evidence and guidance from the literature to avoid the failures and drawbacks cogently criticized in subsequent analyses. As a matter of fact, ANVUR continued to ignore the above recommendations, even for the second VQR.

In 2013, Abramo, Cicero, and D'Angelo (2013) warned that in bibliometric areas, national research assessment exercises based on a subset of publications can be a complete waste of money. To show that, the authors ranked Italian universities by decreasing latitude from north to south, and found it comparable to the VTR ranking. For the individual disciplines, the results actually showed greater accuracy than the VTR in half the disciplines and lesser accuracy in three out of eight. However, when comparing SCImago university rankings by average citation impact (freely accessible on the web), the authors found that these lists outperformed the VTR, both at the university level and by discipline. The authors concluded with a plea: "Governments in general, and especially the Italian government, should question the competencies of those who are planning national evaluation exercises, or at least ensure that there is a sufficient exchange of knowledge between scholars and practitioners to ensure maximum efficiency, effectiveness and fairness in their conduct."

Contrary to the full counting method adopted in the 2004-2010 VQR for assigning contributions to multi-authored publications, in 2013 Abramo, D'Angelo and Rosati (2013a) recommended the adoption of fractional counting in general, and weighted counting in those disciplines where the practice is to signal the contribution of each co-author by its position in the byline. The same authors (2013b) then measured the distortions caused by full counting, when ranking institutions by research productivity for the life sciences. In 2014, as already mentioned, Abramo, D'Angelo and Di Costa (2014) measured the inefficiency of universities in selecting products for submission to the VQR 2004-2014, revealing that what VQR actually measured was not the real performance of institutions. In 2015, Abramo and D'Angelo (2015) identified the 2004-2010 VQR's methodological weaknesses and measured the distortions that resulted from them in the university performance rankings. The results were truly alarming, considering that Italy adopts selective funding of institutions based on VQR rankings:[4] 92% of universities would have a different rank if performance were correctly measured; 69% of universities in Q1 for biology by the VQR would not have placed in Q1; the same statistic was 50% for industrial engineering and 46% for mathematics. The authors reiterated the courteous plea: "the

---

[4] In 2016, about 20% of total public funds to universities were allocated on the basis of 2011-2014 VQR results: http://attiministeriali.miur.it/media/299927/tabella4totale_ffo_2016.pdf. Last accessed 12 May 2017.

Italian case indicates the need for the various national agencies to take more care in the planning stages of assessment exercises, and to call on assistance from the most qualified professionals in the field." It went unheard, or without action. This is shown both in the profile of the new members of ANVUR executive committee, selected by the Minister of Education, University and Research, and the specifications of the 2011-2014 VQR. Although we had vowed to stop wasting time in the offer of unheeded suggestions and recommendations, we could not responsibly refrain from intervening, once again (Abramo & D'Angelo, 2016), when ANVUR announced the magic C-J combination for the next 2011-2014 VQR. We stated: "Notwithstanding the strong criticisms raised in the scientific arena to the methods and indicators used in the 2004-2010 VQR, the 2011-2014 VQR adheres to exactly the same framework, in almost total disregard of both the criticisms and recommendations for improvement" (Abramo & D'Angelo, 2016).

We were not the only ones expressing criticism at the "international" level (Baccini, 2016; Baccini & De Nicolao, 2016). The list of published criticisms goes far beyond the two works acknowledged by BCG&M. The criticism of all authors, including ourselves, was never detached from recommendations. As citizens and civil servants, we are genuinely and patiently motivated by strong willingness to contribute with our knowledge to the betterment of the Italian research evaluation system.

## 5. Conclusions

To conclude, we go back to the original question regarding the failure of knowledge transfer from scholars to practitioners. In order for knowledge transfer to take place, two fundamental conditions are necessary. First, the recipients need to be willing to acquire knowledge from outside. Second, their own stocks and levels of knowledge in the area need to be sufficient for them to understand the new knowledge, and elaborate and incorporate it in operational processes and services. In the Italian case, if we accept that they were able, then the members of the two ANVUR executive committees were clearly never willing to transfer scientometric knowledge from outside. Yet this was exactly what they should have done. Clearly these are highly accomplished individuals, capable of assimilating new knowledge, yet their publication portfolios show no trace of specific knowledge for the roles which they requested. They have certainly undervalued the difficulties of the tasks they took on. It is not possible to shift, in the matter of a moment, from research in telecommunications or tasks as head physician to formulation of a national research assessment exercise. What started as a stroll up a Roman hill has been transformed in a perilous ascent of Everest, and we are afraid these climbers still have not noticed. Their stock and level of knowledge and/or their willingness to take in information were evidently not adequate for self-learning. Doubly sad for all is that, six years on from 2011, it is clear that a properly designed national agency for research evaluation, in the hands of proficient scholars in the field since the beginning, would by now have been a world renowned center of excellence.